\documentclass[fleqn,11pt]{article}

\usepackage{latexsym,ifthen,epsfig,color}
\usepackage{amsmath,amssymb,amsthm}
\usepackage{float}

\newcommand{\join}{\text{\textcircled{{\footnotesize 1}}}}
\newcommand{\cojoin}{\text{\textcircled{{\footnotesize 0}}}}

\newcommand{\NP}{\ensuremath{\mathbb{NP}}}

\newtheorem{theorem}{Theorem}
\newtheorem{lemma}{Lemma}
\newtheorem{corollary}{Corollary}
\newtheorem{proposition}{Proposition}

\newtheorem{clai}{Claim}
\newtheorem{observation}{Observation}

\begin{document}

\title{Dominating Induced Matchings for $P_7$-Free Graphs in Linear Time}

\author{
Andreas Brandst\"adt\footnote{Fachbereich Informatik, 
Universit\"at Rostock, A.-Einstein-Str. 21, D-18051 Rostock, Germany,
{\texttt ab@informatik.uni-rostock.de}}
\and
Raffaele Mosca\footnote{Dipartimento di Scienze, Universit\'a degli Studi ``G. D'Annunzio''
Pescara 65121, Italy.
{\texttt r.mosca@unich.it}}
}

\maketitle

\begin{abstract}
Let $G$ be a finite undirected graph with edge set $E$. An edge set $E' \subseteq E$ is an {\em induced matching} in $G$ if the pairwise distance of the edges of $E'$ in $G$ is at least two;
$E'$ is {\em dominating} in $G$ if every edge $e \in E \setminus E'$ intersects some edge in $E'$. The \emph{Dominating Induced Matching Problem} (\emph{DIM}, for short) asks for the existence of an induced matching $E'$ which is also dominating in $G$; this problem is also known as the \emph{Efficient Edge Domination} Problem. 

The DIM problem is related to parallel resource allocation problems, encoding theory and network routing. It is \NP-complete even for very restricted graph classes such as planar bipartite graphs with maximum degree three.
However, its complexity was open for $P_k$-free graphs for any $k \ge 5$; $P_k$ denotes a chordless path with $k$ vertices and $k-1$ edges. We show in this paper that the weighted DIM problem is solvable in linear time for $P_7$-free graphs in a robust way. 
\end{abstract}


\section{Introduction}\label{sec:intro}

Let $G$ be a simple undirected graph with vertex set $V$ and edge set $E$. A subset $M$ of $E$ is an {\em induced matching} in $G$ if the $G$-distance of every pair of edges $e,e' \in M$, $e \neq e'$, is at least two, i.e., $e \cap e' = \emptyset$ and there is no edge $xy \in E$ with $x \in e$ and $y \in e'$.
A subset $M \subseteq E$ is a {\em dominating edge set} if every edge $e \in E \setminus M$ shares an endpoint with some edge $e' \in M$, i.e., if $e \cap e' \not= \emptyset$. A {\em dominating induced matching} ({\em d.i.m.} for short) is an induced matching which is also a dominating edge set.

Let us say that {\em an edge $e \in E$ is matched by $M$} if $e \in M$ or there is $e' \in M$ with $e \cap e' \neq \emptyset$. Thus, $M$ is a d.i.m. of $G$ if and only if every edge of $G$ is matched by $M$ but no edge is matched twice. 

The {\em Dominating Induced Matching Problem} ({\em DIM}, for short) asks whether a given graph has a dominating induced matching. This can also be seen as a special 3-colorability problem, namely the partition into three independent vertex sets $A,B,C$ such that $G[B \cup C]$ is an induced matching: If $M \subseteq E$ is a d.i.m. of $G$ then the vertex set has the partition $V= A \cup V(M)$ with independent vertex set $A$, and independent sets $B,C$ with $B \cup C=V(M)$.

Dominating induced matchings are also called {\em edge packings} in some papers, and DIM is known as the {\em Efficient Edge Domination Problem} ({\em EED} for short). A brief history of EED as well as some applications in the fields of resource allocation, encoding theory and network routing are presented in \cite{GriSlaSheHol1993} and \cite{LivSto1988}.

Grinstead et al. \cite{GriSlaSheHol1993} show that EED is \NP-complete in general.
It remains hard for bipartite graphs \cite{LuTan1997}. In particular, \cite{LuKoTan2002} shows the intractability of EED for planar bipartite graphs and \cite{CarLoz2008} for very restricted bipartite graphs with maximum degree three (the restrictions are some forbidden subgraphs). In \cite{BraHunNev2010}, it is shown that the problem remains \NP-complete for planar bipartite graphs with maximum degree three but is solvable in polynomial time for hole-free graphs (which was an open problem in \cite{LuKoTan2002} and is still mentioned as an open problem in \cite{CarKorLoz2011}; actually, \cite{CarKorLoz2011,LuKoTan2002} mention that the complexity of DIM is an open problem for weakly chordal graphs which are a subclass of hole-free graphs). 

In \cite{CarKorLoz2011}, as another open problem, it is mentioned that for any $k \ge 5$, the complexity of DIM is unknown for the class of $P_k$-free graphs. Note that the complexity of the related problems Maximum Independent Set and Maximum Induced Matching is unknown for $P_5$-free graphs, and a lot of work has been done on subclasses of $P_5$-free graphs.     

In this paper, we show that for $P_7$-free graphs, DIM is solvable in linear time. Actually, we consider the edge-weighted optimization version of DIM, namely the {\em Minimum Dominating Induced Matching Problem} ({\em MDIM}), which asks for a dominating induced matching $M$ in $G=(V,E)$ of minimum weight with respect to some given weight function $\omega: E \rightarrow \mathbb{R}$ (if existent).   

For $P_5$-free graphs, DIM is solvable in time ${\cal O}(n^2)$ as a consequence of the fact that the clique-width of ($P_5$,gem)-free graphs is bounded \cite{BraKra2005,BraLeMos2005} and a clique-width expression can be constructed in time ${\cal O}(n^2)$ \cite{BodBraKraRaoSpi2005}. In \cite{CarKorLoz2011}, it is mentioned that DIM is expressible in a certain kind of Monadic Second Order Logic, and in \cite{CouMakRot2000}, it was shown that such problems can be solved in linear time on any class of bounded clique-width assuming that the clique-width expressions are given or can be determined in the same time bound. %In a separate note \cite{BraMos2011} we give a linear time algorithm for weighted DIM on $P_5$-free graphs.

It is well known that the clique-width of cographs (i.e., $P_4$-free graphs) is at most two (and such clique-width expressions can be determined in linear time) and thus the DIM problem can be solved in linear time on cographs. In section \ref{sec:cographs} we give a simple characterization of cographs having a d.i.m.

Our algorithm for $P_7$-free graphs is based on a structural analysis of such graphs having a d.i.m. It is robust in the sense of \cite{Spinr2003} since it is not required that the input graph is $P_7$-free; our algorithm either determines an optimal d.i.m. correctly or finds out that $G$ has no d.i.m. or is not $P_7$-free. 

\section{Further Basic Notions}\label{sec:basicnotions}

Let $G$ be a finite undirected graph without loops and multiple edges. Let $V$ denote its vertex set and $E$ its edge set; let $|V|=n$ and $|E|=m$.
For $v \in V$, let $N(v):=\{u \in V \mid uv \in E\}$ denote the {\em open neighborhood of $v$}, and let $N[v]:=N(v) \cup \{v\}$ denote the {\em closed neighborhood of $v$}. If $xy \in E$, we also say that $x$ and $y$ {\em see each other}, and if  $xy \not\in E$, we say that $x$ and $y$ {\em miss each other}. A vertex set $S$ is {\em independent} (or {\em stable}) in $G$ if for every pair of vertices $x,y \in S$, $xy \not\in E$. A vertex set is a {\em clique} in $G$ if for every pair of vertices $x,y \in S$, $x \neq y$, $xy \in E$ holds. For $uv \in E$ let $N(uv):= N(u) \cup N(v) \setminus \{u,v\}$ and $N[uv]:= N[u] \cup N[v]$.
Distinct vertices $x$ and $y$ are {\em true twins} if $N[x]=N[y]$. 

For $U \subseteq V$, let $G[U]$ denote the induced subgraph of $G$ with vertex set $U$, hence, the graph which contains exactly the edges $xy \in E$ with both vertices $x$ and $y$ in $U$.

Let $\overline{G}$ (or co-$G$) denote the {\em complement graph} of $G=(V,E)$, i.e., $\overline{G}=(V,\overline{E})$ with $xy \in \overline{E}$ if and only if $x \neq y$ and $xy \not\in E$. 

Let $A$ and $B$ be disjoint vertex sets in $G$. If every vertex from $A$ sees (misses, respectively) every vertex from $B$, we denote this by $A \join B$ (by $A \cojoin B$, respectively). 

A set $H$ of at least two vertices of a graph $G$ is called \emph{homogeneous} if $H \not= V(G)$ and every vertex outside $H$ is adjacent to all vertices in $H$ or to no vertex in $H$. Obviously, $H$ is homogeneous in $G$ if and only if $H$ is homogeneous in the complement graph $\overline{G}$. 

A homogeneous set $H$ is \emph{maximal} if no other homogeneous set properly contains $H$. It is well known that in a connected graph $G$ with 
connected complement $\overline{G}$, the maximal homogeneous sets are pairwise disjoint and can be determined in linear time (see, e.g., \cite{McCSpi1999}). 

\begin{figure}
  \begin{center}
    \epsfig{file=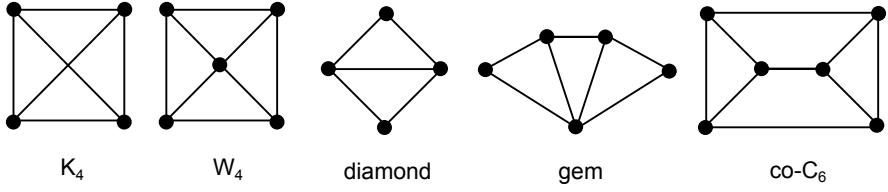}
    \caption{$K_4$, $W_4$, diamond, gem, and co-$C_6$.}
    \label{K4gem}
  \end{center}
\end{figure}

A {\em chordless path} $P_k$ ({\em chordless cycle} $C_k$, respectively) has $k$ vertices, say $v_1,\ldots,v_k$, and edges $v_iv_{i+1}$, $1 \le i \le k-1$ (and $v_kv_1$, respectively). We say that such a path (cycle, respectively) has length $k$.
Let $K_i$ denote the clique with $i$ vertices. Let $K_4-e$ or {\em diamond} be the graph with four vertices and five edges, say vertices $a,b,c,d$ and edges $ab,ac,bc,bd,cd$; its {\em mid-edge} is the edge $bc$.
Let $W_4$ denote the graph with five vertices consisting of a $C_4$ and a universal vertex (see Figure \ref{K4gem}). Let $K_{1,k}$ denote the star with one universal vertex and $k$ independent vertices. A star is {\em nontrivial} if it contains a $P_3$ or an edge, otherwise it is {\em trivial}. 

For two vertices $x,y \in V$, let $dist_G(x,y)$ denote the {\em distance between $x$ and $y$ in $G$}, i.e., the length of a shortest path between $x$ and $y$ in $G$.
The {\em distance of two edges} $e,e' \in E$ is the length of a shortest path between $e$ and $e'$, i.e., $dist_G(e,e')= \min\{dist_G(u,v) \mid u \in e, v \in e'\}$.
In particular, this means that $dist_G(e,e')=0$ if and only if $e \cap e' \neq \emptyset$.
For a vertex $x$, let $N_i(x)$ denote the {\em distance levels of $x$}: $N_i(x):=\{v \mid dist_G(v,x) = i\}$. Thus, $N_1(x)=N(x)$.
For an edge $xy$, let $N_i(xy)$ denote the {\em distance levels of $xy$}: $N_i(xy):=\{z \mid dist_G(z,xy) = i\}$. Thus, $N_1(xy)=N(xy)$.

A {\em connected component} of $G$ is a maximal vertex subset $U \subseteq V$ such that all pairs of vertices of $U$ are connected by paths in $G[U]$.
A {\em $2$-connected component} of $G$ is a maximal vertex subset $U \subseteq V$ such that all pairs of vertices of $U$ are connected by at least two vertex-disjoint paths in $G[U]$. The 2-connected components are also called {\em blocks}. It is well known that the blocks of a graph can be determined in linear time \cite{HopTar1973} (see also \cite{AhoHopUll1974}).

For a set ${\cal F}$ of graphs, a graph $G$ is called {\em ${\cal F}$-free} if $G$ contains no induced subgraph from ${\cal F}$.
A {\em hole} is a $C_k$ for some $k \ge 5$. A graph is {\em hole-free} if it is $C_k$-free for all $k \ge 5$.
A graph is {\em chordal} if it is $C_k$-free for all $k \ge 4$.
A graph is {\em weakly chordal} if it is $C_k$-free and $\overline{C_k}$-free for all $k \ge 5$.

If $M$ is a d.i.m., an edge is {\em matched by $M$} if it is either in $M$ or shares a vertex with some edge in $M$. Likewise, a vertex is {\em matched} if it is in $V(M)$.

Note that $M$ is a d.i.m. in $G$ if and only if it is a dominating vertex set in the line graph $L(G)$ and an independent vertex set in the square $L(G)^2$. Thus, the DIM problem is simultaneously a packing and a covering problem.  

\section{Simple Properties of Graphs With Dominating Induced Matching}\label{sec:properties}

The following observations are helpful (some of them are mentioned e.g. in \cite{BraHunNev2010}):

\begin{observation}\label{dimC3C5C7C4}
Let $M$ be a d.i.m. in $G$.
\begin{itemize}
\item[$(i)$] $M$ contains at least one edge of every odd cycle $C_{2k+1}$ in $G$, $k \ge 1$,
and exactly one edge of every odd cycle $C_3$, $C_5$, $C_7$ of $G$.
\item[$(ii)$] No edge of any $C_4$ can be in $M$.
\item[$(iii)$] If $C$ is a $C_6$ then either exactly two or none of the $C$-edges are in $M$.
\end{itemize}
\end{observation}

\noindent
{\bf Proof.}
(i): Let $C$ be an odd cycle $C_{2k+1}$ in $G$, $k \ge 1$, with vertices $v_1,\ldots,v_{2k+1}$ and edges $v_iv_{i+1}$, $i \in \{1,\ldots,2k+1\}$ (index arithmetic modulo $2k+1$). Suppose first that none of the edges of $C$ are in $M$. Then the edge $v_1v_2$ must be matched by an $M$-edge, say by $v_1x$, $x \neq v_2,v_{2k+1}$. Now the edge $v_2v_3$ must be matched in $v_3$ and so on, until finally the edge $v_{2k}v_{2k+1}$ must be matched in $v_{2k+1}$ but now two $M$-edges are in distance one - contradiction.

Now for $C_3$'s and $C_5$'s in $G$, obviously not more than one edge can be in $M$. If for a $C_7$, two edges would be in $M$, say $v_1v_2 \in M$ and $v_4v_5 \in M$ then $v_6v_7$ cannot be matched - contradiction.

\noindent
(ii): If $(v_1,v_2,v_3,v_4)$ is a $C_4$ in $G$ then if $v_1v_2 \in M$, $v_3v_4$ is not matchable.

\noindent
(iii): This condition obviously holds.
\qed \medskip

If an edge $e \in E$ is contained in any d.i.m. of $G$, we call it {\em mandatory} (or {\em forced}) in $G$. Mandatory edges are useful for some kinds of reductions.

\begin{observation}\label{obs:diamondbutterfly}
The mid-edge of any diamond in $G$ is mandatory.
\end{observation}

If an edge $xy$ is mandatory, we can reduce the graph as follows: Delete $x$ and $y$ and all edges incident to $x$ and $y$, and give all edges in distance one to $xy$ the weight $\infty$. This means that these edges are not in any d.i.m. of finite weight in $G$. For a set $M$ of mandatory edges, let {\em Reduced$(G,M)$} denote the reduced graph as defined above. Obviously, this graph is an induced subgraph of $G$ and can be determined in linear time for given $G$ and $M$. Moreover:
\begin{observation}\label{obs:reducedDIM}
Let $M'$ be an induced matching which is a set of mandatory edges in $G$. Then $G$ has a d.i.m. $M$ if and only if Reduced$(G,M')$ has a d.i.m. $M \setminus M'$.
\end{observation}
We can also color red all vertices in distance one to a mandatory edge; subsequently, edges $ab$ with a red vertex $a$ cannot be matched in vertex $a$; they have to be matched in vertex $b$. If also $b$ is red then $G$ has no d.i.m. 

\medskip

Subsequently, as a kind of preprocessing, some of the mid-edges of diamonds will be determined. Since it would be too time-consuming to determine all diamonds in $G$, we will mainly find such diamonds whose mid-edges are edges between true twins having at least two common neighbors. These are contained in maximal homogeneous sets which can be found in linear time. 

\medskip

Since the edges of any d.i.m. must have pairwise distance at least two, we obtain:
\begin{observation}\label{obs:dimexneighbhoods}
If $G$ has a d.i.m. then for all vertices $v$, $G[N(v)]$ is the disjoint union of at most one star with $P_3$, and of edges and vertices.
\end{observation}

\noindent
{\bf Proof.}
Let $G$ have a d.i.m. $M$. Then by Observation \ref{dimC3C5C7C4} (i), $M$ contains an edge of every triangle, and by Observation \ref{obs:diamondbutterfly}, any $P_3$ $abc$ in $N(v)$ generates a mandatory edge $bv$. Thus, if $N(v)$ contains a cycle or $P_4$, we obtain a contradiction to the distance requirements in $M$. This means that $N(v)$ is a $P_4$-free forest, i.e., a disjoint union of stars. If there are two stars with $P_3$ in $N(v)$ then again we obtain a contradiction.
\qed \medskip

From the previous observations, it follows (see Figure \ref{K4gem} for $K_4,W_4$, gem, and  $\overline{C_6}$):
\begin{corollary}\label{cly:k4gemfree}
If $G$ has a d.i.m. then $G$ is $K_4$-free, $W_4$-free, gem-free and $\overline{C_k}$-free for any $k \ge 6$.
\end{corollary}

Now we deal with homogeneous sets in $G$. 

\begin{corollary}\label{homsets}
Let $G$ have a d.i.m. and let $H$ be a homogeneous set in $G$. 
\begin{itemize}
\item[$(i)$] If $H$ contains an edge then $N(H)$ is stable. 
\item[$(ii)$] If $|N(H)| \ge 2$ then $H$ is either a stable set or a disjoint union of edges. 
\item[$(iii)$] Vertices $x$ and $y$ are true twins with at least two common neighbors in $G$ if and only if they appear as an edge in a homogeneous set $H$ with $|N(H)| \ge 2$.
\end{itemize}
\end{corollary}

\noindent
{\bf Proof.} Let $G$ have a d.i.m. and let $H$ be a homogeneous set in $G$.
\noindent
(i): If $H$ contains an edge then since $G$ is $K_4$-free, $N(H)$ is stable. 

\noindent
(ii): If $|N(H)| \ge 2$ then by Observation \ref{obs:dimexneighbhoods} and Corollary \ref{cly:k4gemfree}, $H$ must be $P_3$-free, i.e., is a disjoint union of cliques. Since $G$ is $K_4$-free, the cliques are edges or vertices. If there is an edge $uv$ in $H$ and there is a component $w \in H$ consisting of a single vertex then obviously, $uv \in M$ and for any $a \in N(H)$, the edge $aw$ cannot be matched - contradiction. 

\noindent
(iii): If $x$ and $y$ are true twins then $x,y$ are contained in a (maximal) homogeneous set. On the other hand, if $x$ and $y$ with $xy \in E$ appear in a $P_3$-free homogeneous set then $x$ and $y$ are true twins. 
\qed

\medskip

The following procedure uses Observation \ref{obs:dimexneighbhoods} and the fact that for a homogeneous set $H$ with $|N(H)|=1$, say $N(H)=\{z\}$, all connected components of $H$ together with $z$ are leaf blocks in $G$. 

\medskip

\noindent
{\bf Procedure Hom-1-DIM:}

\medskip

\noindent
{\bf Given:} A homogeneous set $H$ in $G$ with $N(H)=\{z\}$.  

\noindent
{\bf Task:} Determine some mandatory edges or find out that $G$ has no d.i.m. 

\begin{enumerate}
\item[(a)] If $H$ contains a cycle or $P_4$ then STOP - $G$ has no d.i.m. 

\item[(b)] ({\em Now $H$ is a $P_4$-free forest.}) If $H$ contains at least two $P_3$'s then STOP - $G$ has no d.i.m. 

\item[(c)] ({\em Now $H$ is a $P_4$-free forest which contains at most one $P_3$.}) If $H$ contains exactly one $P_3$, say $abc$ then $M:=M \cup \{bz\}$. If another connected component of $H$ contains an edge then STOP - $G$ has no d.i.m.    

\item[(d)] ({\em Now $H$ is a $P_3$-free forest, i.e., a disjoint union of edges $E'(H)$ and vertices $V'(H)$.}) If $E'(H)$ contains at least two edges then $M:=M \cup E'(H)$. If $V'(H) \neq \emptyset$ then STOP - $G$ has no d.i.m. 

\item[(e)] ({\em Now $H$ is a disjoint union of at most one edge and vertices $V'(H)$.}) If there is an edge $ab$ in $H$ and $V'(H) \neq \emptyset$ then $M:=M \cup \{az\}$ or $M:=M \cup \{bz\}$ (depending on the better weight).   
\end{enumerate}

We postpone the discussion of the two final cases $E'(H)=\{ab\}$ and $V'(H)=\emptyset$ or $E'(H)=\emptyset$ and $V'(H)\neq \emptyset$.
Obviously, the following holds:
\begin{lemma}\label{Hom1DIMcorrlintime}
Procedure Hom-1-DIM is correct and can be carried out in linear time. 
\end{lemma}

In the final case of a homogeneous set $H$ with only one neighbor $z$ where $H$ consists of just one edge $ab$, $abz$ forms a leaf block. For graph $G$, let $G^*$ denote the graph obtained from $G$ by omitting all such triangle leaf blocks. Obviously, $G^*$ can be constructed in linear time. We will need this construction in our algorithm $P_7$-Free-DIM for DIM in section \ref{generalP7freecase}.
There, we also need the following transformation: For every triangle leaf block $abc$ with cut-vertex $c$ and corresponding edge weights $w(ab)$, $w(ac)$, $w(bc)$, let $Tr(G,abc)$ be the graph with the same cut-vertex $c$ where the triangle is replaced by a path $a'b'c$ with weights $w(ab)$ for edge $a'b'$ and $\min(w(ac),w(bc))$ for edge $b'c$. Let $Tr(G)$ be the result of applying $Tr(G,abc)$ to all triangle leaf blocks $abc$ of $G$. Obviously, $G$ has a d.i.m. if and only if $Tr(G,abc)$ has a d.i.m., and the optimal weights of d.i.m.'s in $G$ and  $Tr(G,abc)$ are the same. The only problem is the fact that the new graph is not necessarily $P_7$-free when $G$ is $P_7$-free. We will apply this construction only in one case, namely when the internal blocks of $G$ form a distance-hereditary bipartite graph; then $Tr(G)$ is also distance hereditary bipartite.     

\medskip

Finally we need the following:
\begin{proposition}\label{lintimedimchecking}
For a given set $E'$ of edges, it can be tested in linear time whether $E'$ is a d.i.m., and likewise, whether $E'$ is an induced matching.
\end{proposition}

\noindent
{\bf Proof.}
For $E' \subseteq E$, in an array of all vertices in $V$, count the number $m(x)$ of appearances of each vertex of $V$ in the edges of $E'$ by going through all edges in $E'$ once.
\begin{enumerate}
\item Two edges of $E'$ intersect if and only if one of the vertices appears in more than one edge, i.e., if there is a vertex $x$ with $m(x) \ge 2$.  
\item Two edges of $E'$ have distance one if and only if for an edge $xy \in E \setminus E'$, both $m(x) \ge 1$ and $m(y) \ge 1$.
\item $E'$ is dominating if and only if for each edge $xy \in E$, $m(x) \ge 1$ or $m(y) \ge 1$. 
\end{enumerate}
Obviously this can be checked in time ${\cal O}(n+m)$. The first two steps are the test whether $E'$ is an induced matching.
\qed

\section{DIM for Cographs}\label{sec:cographs}

It is well known that a graph is a cograph if and only if its clique-width is at most two. Thus, for solving the DIM problem on cographs, one could use the clique-width argument. However, we give a simple direct way. Obviously, the following holds:
\begin{corollary}\label{cor:coconnwithdim}
If $G$ has a d.i.m. and $\overline{G}$ is not connected then $G$ is a cograph. 
\end{corollary}

For the subsequent characterization of cographs, i.e., $P_4$-free graphs, with d.i.m., we need the following notion:

$G$ is a {\em super-star} if $G$ contains a universal vertex $u$ such that $G[V \setminus \{u\}]$ is the disjoint union of a star and a stable set.
Note that every super-star has a d.i.m. $M$, namely if the star contains a $P_3$ with central vertex $c$ then $M$ consists of the single edge $uc$, and if the star consists of only one edge $ab$, then $\{ua\}$ and $\{ub\}$ are both d.i.m.'s, and the choice of an optimal d.i.m. depends on the weights. If there is no edge in $G[V \setminus \{u\}]$ then any edge $uv$ is a d.i.m., and the choice of an optimal d.i.m. depends on the weights.

For cographs having a d.i.m., there is the following simple characterization:

\begin{proposition}\label{dimcographs}
A connected cograph $G$ has a d.i.m. if and only if it is either a super-star or the join $G=G_1 \join G_2$ of a disjoint union of edges $G_1$ and a stable set $G_2$.
\end{proposition}

\noindent
{\bf Proof.}
Let $G$ be a connected cograph with a d.i.m. $M$. Then, since $G$ is $K_4$-free, $G=G_1 \join G_2$ for some triangle-free (i.e., bipartite) subgraphs $G_1$ and $G_2$.

\medskip

\noindent
{\bf Case 1.} $G_1$ (or $G_2$) contains only one vertex; without loss of generality say $V(G_1)=\{u\}$.

Then by Observation \ref{obs:dimexneighbhoods}, $G_2$ is a disjoint union of at most one star with $P_3$, of edges and vertices. 
If exactly one of the connected components of $G_2$ contains a $P_3$ then this component is a star, say with central vertex $c$, and $uc \in M$. Now the other components of $G_2$ must be single vertices since in every triangle, exactly one edge is in $M$. This shows that in this case, $G$ is a super-star, and an optimal d.i.m. can be chosen as described above. 

If none of these connected components contain $P_3$ then the connected components of $G_2$ are edges and vertices. If at least two such edges exist then all the connected components are edges, otherwise there is no d.i.m. This corresponds to the second case in Proposition \ref{dimcographs}.

If exactly one of the connected components is an edge, say $ab$, and all the others are vertices then $ua$ and $ub$ are possible d.i.m.'s. This is again a special super-star. If there is no edge in $G_2$ then $G$ is simply a star.

\medskip

\noindent
{\bf Case 2.} $G_1$ and $G_2$ contain at least two vertices.

If none of $G_1$, $G_2$ contains an edge then if both $G_1$ and $G_2$ contain at least two vertices, every edge is in a $C_4$ and therefore not in $M$ - contradiction.

If $G_1$ contains an edge then by Corollary \ref{homsets} (i), $G_2$ is edgeless and by Corollary \ref{homsets} (ii), $G_1$ is a disjoint union of edges. In this case, the uniquely determined d.i.m. of $G$ is the set of edges in $G_1$.

Conversely, it is easy to see that any super-star has a d.i.m., and likewise any join of a disjoint union of edges and a stable set has a d.i.m.
\qed \medskip

\begin{corollary}\label{dimcographslinrec}
Cographs with d.i.m. can be recognized in linear time.
\end{corollary}

The following uses Proposition \ref{dimcographs}:

\medskip

\noindent
{\bf Procedure Cograph-DIM:} 

\medskip

\noindent
{\bf Given:} A connected cograph $G$.  

\noindent
{\bf Task:} Decide whether $G$ has a d.i.m. and if yes, determine a d.i.m. of $G$.   

\begin{enumerate}
\item[(a)] Check whether $G$ is either a super-star or the join of a disjoint union of edges and a stable set. If yes then $G$ has a d.i.m. as described above, and if not then STOP - $G$ has no d.i.m. 
\end{enumerate}

\section{Structure of $P_7$-free Graphs With Dominating Induced Matching}\label{P7free}

Throughout this section, let $G=(V,E)$ be a connected $P_7$-free graph having a d.i.m. Recall that if $M$ is a d.i.m. of $G$ then the vertex set $V$ has the partition $V= I \cup V(M)$ with independent vertex set $I$. We suppose that $xy \in M$ is an edge in a $P_3$ and consider the distance levels $N_i=N_i(xy)$, $i \ge 1$, with respect to the edge $xy$. Note that every edge of an odd hole $C_5$, $C_7$, respectively, is in a $P_3$. For triangles $abc$, this is not fulfilled if $a$ and $b$ are true twins. However, true twins with at least two common neighbors will lead to mandatory edges, and true twins $a,b$ with only one common neighbor $c$ form a leaf block $abc$ which will be temporarily omitted by constructing $G^*$ and looking for an odd cycle in $G^*$.  

\subsection{Distance levels with respect to an $M$-edge}\label{levelstructure}

Since we assume that $xy \in M$, clearly, $N_1 \subseteq I$ and thus:
\begin{equation}\label{NxysubI}
N_1 \mbox{ is a stable set.}
\end{equation}

Moreover, no edge between $N_1$ and $N_2$ is in $M$. Since $N_1 \subseteq I$ and all neighbors of vertices in $I$ are in $V(M)$, we have:
\begin{equation}\label{N2M2U2}
N_2 \mbox{ is the disjoint union of edges and vertices in } M.
\end{equation}

Let $M_2$ denote the set of edges in $N_2$ and let $S_2$ denote the set of isolated vertices in $N_2$; $N_2=V(M_2) \cup S_2$. Obviously:
\begin{equation}\label{M2subM}
M_2 \subseteq M \mbox{ and } S_2 \subseteq V(M).
\end{equation}

Let $M_3$ denote the set of $M$-edges with one endpoint in $S_2$ (and the other endpoint in $N_3$).

\medskip

Since $xy$ is contained in a $P_3$, i.e., there is a vertex $r$ such that $y,x,r$ induce a $P_3$, we obtain some further properties: 
\begin{equation}\label{N5empty}
N_5=\emptyset.
\end{equation}

\noindent
{\em Proof of} (\ref{N5empty}):
If there is a vertex $v_5 \in N_5$ then there is a shortest path $(v_5$, $v_4$, $v_3$, $v_2$, $v_1)$, $v_i \in N_i$, $i =1,\ldots,5$, connecting $v_5$ and a neighbor $v_1$ of $x$ or $y$. If $v_2r \in E$ then $v_5,v_4,v_3,v_2,r,x,y$ is a $P_7$, and if $v_2$ is nonadjacent to any personal neighbor of $x$ with respect to $y$ then $v_5,v_4,v_3,v_2,v_1,x,r$ is a $P_7$ or $v_5,v_4,v_3,v_2,v_1,y,x$ is a $P_7$ - a contradiction which shows (\ref{N5empty}). $\diamond$

\medskip

This kind of argument will be used later again - we will say that the subgraph induced by $x,y,N_1,v_2,v_3,v_4,v_5$ contains an induced $P_7$.

Obviously, by (\ref{M2subM}) and the distance condition, the following holds:
\begin{equation}\label{edgesN3N4}
\mbox{ No edge in } N_3 \mbox{ and no edge between } N_3 \mbox{ and } N_4 \mbox{ is in } M.
\end{equation}

\medskip

Furthermore the following statement holds.

\begin{equation}\label{N4M4U4}
N_4 \mbox{ is the disjoint union of edges and vertices.}
\end{equation}

\noindent
{\em Proof of} (\ref{N4M4U4}):
The proof is very similar to the one of (\ref{N5empty}): Let $uv$ be an edge in $N_4$ and let $w \in N_3$ see $u$; then $w$ must see also $v$ since $G$ is $P_7$-free (recall the existence of $r$ in a $P_3$ with $x$ and $y$). Then $N_4$ must be $P_3$-free - otherwise any neighbor $w \in N_3$ of a $P_3$ $abc$ in $N_4$ would induce a diamond $w,a,b,c$ and then edge $wb$ is mandatory in contradiction to Observation \ref{obs:diamondbutterfly} and condition (\ref{edgesN3N4}). Moreover, $N_4$ is triangle-free (otherwise there is a $K_4$ in contradiction to Corollary \ref{cly:k4gemfree}).
Then $N_4$ is a disjoint union of edges and vertices which shows (\ref{N4M4U4}). $\diamond$

\medskip

Let $M_4$ denote the set of edges in $N_4$ and let $S_4$ denote the set of isolated vertices in $N_4$; $N_4 = V(M_4) \cup S_4$. Note that by (\ref{N5empty}) and (\ref{edgesN3N4}), $S_4 \subseteq I$.

Since every edge $ab$ in $N_4$ together with a predecessor $c$ in $N_3$ forms a triangle, and $ac, bc \notin M$, by (\ref{edgesN3N4}) necessarily:
\begin{equation}\label{edgesM4M}
M_4 \subseteq M.
\end{equation}

By Observation \ref{dimC3C5C7C4} (i), in every odd cycle $C_3$, $C_5$ and $C_7$ of $G$, exactly one edge must be in $M$. Thus, (\ref{edgesN3N4}) implies:
\begin{equation}\label{N3S4bip}
N_3 \cup S_4 \mbox{ is bipartite.}
\end{equation}

Note that in general, $N_3$ is not a stable set.

\medskip

Let $T_{one} := \{t \in N_3: |N(t) \cap S_2| = 1\}$, and $T_{two} := \{t \in N_3: |N(t) \cap S_2| \geq 2\}$. Note that if $uv$ is an edge with $u \in T_{two}$ then $uv \not\in M$ and $uv$ must be matched by an $M$-edge at $v$ since it cannot be matched at $u$ because of the distance condition; in particular, $T_{two} \subseteq I$.

In general, (\ref{edgesN3N4}) will lead to some forcing conditions since the edges in $N_3$ and between $N_3$ and $N_4$ have to be matched. If an edge $uv \in E$ cannot be matched at $u$ then it has to be matched at $v$ -  in this case, as described later, we color the vertex $v$ green if it has to be matched by an $M_3$ edge. (For an algorithm checking the existence of a d.i.m., it is useful to observe that if vertices in distance one get color green then no d.i.m. exists.)

Let $S_3:= (N(M_2) \cap N_3) \cup (N(M_4) \cap N_3) \cup T_{two}$. Then $S_3 \subseteq N_3$ and $S_3 \subseteq I$. Furthermore, since $S_4 \subseteq I$, one obtains:
\begin{equation}\label{S3S4stable}
S_3 \cup S_4 \mbox{ is a stable set.}
\end{equation}

Let $T^*_{one} := T_{one} \setminus S_3$. Then $N_3=S_3 \cup T^*_{one}$ is a partition of $N_3$. In particular, $T^*_{one}$ contains the $M$-mates of the vertices of $S_2$. Recall that $M_3$ denotes the set of $M$-edges with one endpoint in $S_2$ (and the other endpoint in $T^*_{one}$).

\subsection{Edges in and between $T_i$ and $T_j$, $i \neq j$}\label{TiTjedges}

Let $S_2 = \{u_1,u_2, \ldots,u_k\}$, and let $T_i := T_{one}^{*} \cap N(u_i)$, $i = 1,\ldots,k$. Then $T_{one}^{*}=T_1 \cup \ldots \cup T_k$ is a partition of $T_{one}^{*}$. The following condition is necessary for the existence of $M_3$:
\begin{equation}\label{Tinonempty}
\mbox{For all } i = 1,\ldots,k, T_i \neq \emptyset, \mbox{ and exactly one vertex of } T_i \mbox{ is in } V(M_3).
\end{equation}

Recall that by Observation \ref{obs:dimexneighbhoods}, $G[T_i]$ is the disjoint union of at most one star with $P_3$, and of edges and vertices. Furthermore, $G[T_i]$ cannot contain two edges, i.e., the following statement holds for all $i = 1,\ldots,k$:
\begin{equation}\label{conncompTi}
G[T_i] \mbox{ is a disjoint union of vertices and at most one star $Y_i$ with an edge.}
\end{equation}

\noindent
{\em Proof of} (\ref{conncompTi}): Assume that there are two edges, say $ab$ and $a'b'$, in $T_i$. Then in both triangles $u_iab$, $u_ia'b'$, exactly one edge has to be in $M$ but both contain $u_i$ - contradiction.
$\diamond$ \medskip

Assume that $T_i$ contains the star $Y_i$ with an edge. 
\begin{equation}\label{isolated-ei}
\mbox{For all } i,j = 1,\ldots,k, i \neq j, Y_i \mbox{ sees no vertex of } T_j.
\end{equation}

\noindent
{\em Proof of} (\ref{isolated-ei}): Let $t'_it''_i$ be an edge of $Y_i$. By contradiction assume that a vertex $t_j \in T_j$, $i \neq j$, is adjacent to $Y_i$, say $t_j$ sees $t''_i$. Then, since by (\ref{N3S4bip}), $G[T^*_{one}]$ is triangle-free, $t_j$ is nonadjacent to $t'_i$, and now $x,y,N_1,u_j,t_j,t''_i,t'_i$ induce a subgraph of $G$ containing a $P_7$.
$\diamond$ \medskip

\begin{clai}\label{edgesbetweenTiTj}
For all $i = 1,\ldots,k$, there is at most one $j \neq i$ such that a vertex in $T_i$ sees a vertex in $T_j$.
\end{clai}

\noindent
{\em Proof of Claim} \ref{edgesbetweenTiTj}: By contradiction assume that there are two indices $j \neq h$ such that some vertices in $T_i$ see vertices in $T_j$ and $T_h$.

\noindent
{\em Case $1$.} If there is a vertex $t_i \in T_i$ which sees a vertex $t_j \in T_j$ and $t_h \in T_h$ then, since there is no triangle in $N_3$, $t_j$ misses $t_h$, and then $x,y,N_1,u_h,t_h,t_j,t_i$ induce a subgraph of $G$ containing a $P_7$
(recall the existence of a $P_3$ with $x,y$ and vertex $r \in N_1$).

\noindent
{\em Case $2$.} Thus, assume that there are two vertices $t'_i,t''_i \in T_i$ such that $t'_i$ sees a vertex $t_j \in T_j$ and $t''_i$ sees a vertex $t_h \in T_h$. Clearly, by (\ref{isolated-ei}), $t'_it''_i \notin E$, and by Case 1, $t'_it_h \notin E$, $t''_it_j \notin E$. Moreover, $t_jt_h \notin E$, otherwise we are in Case 1 again. Now $u_j,t_j,t'_i,u_i,t''_i,t_h,u_h$ induce a $P_7$ - contradiction.
$\diamond$ \medskip

Let us say that {\em $T_i$ sees $T_j$} if there are vertices in $T_i$ and $T_j$ which see each other. Now by Claim \ref{edgesbetweenTiTj}, for every  $i = 1,\ldots,k$, $T_i$ either sees no $T_j$, $j \neq i$, and in this case let us say that {\em $T_i$ is isolated}, or sees exactly one
$T_j$, $j \neq i$, in which case we say that {\em $T_i$ and $T_j$ are paired}.

\begin{clai}\label{eiTiTj}
If $T_i$ and $T_j$ are paired then $G[T_i \cup T_j]$ contains at most two components among the four following ones: $Y_i$ $($defined above$)$, $Y_j$ $($defined above$)$, $Y'_i$ which is a star with center in $T_i$ and the other vertices in $T_j$, $Y'_j$ which is a star with center in $T_j$ and the other vertices in $T_i$; in particular, at most one from $\{Y_i,Y_j\}$ does exist.
\end{clai}

\noindent
{\em Proof of Claim} \ref{eiTiTj}:
By (\ref{Tinonempty}) and since each edge of $G$ must be matched by $M$, $G[T_i \cup T_j]$ contains at most two components. By (\ref{conncompTi}) and (\ref{isolated-ei}) it is enough to focus on the possible components of $G[T_i \cup T_j]$ with vertices in both $T_i$ and $T_j$. In particular, by (\ref{conncompTi}) each such component is a star with center in $T_i$ (in $T_j$, respectively) and the other vertices in $T_j$ (in $T_i$, respectively); if any of such stars contains a $P_3$ then its center $c$ belongs to $V(M_3)$ (in fact otherwise, $c$ would have have two neighbors in $T_i$ or in $T_j$, and such neighbors should belong to $V(M)$, a contradiction to (\ref{Tinonempty})); then if such stars exist and contain $P_3$, their centers belong to $T_i$ and $T_j$ respectively; then one obtains the stars described in the claim. Finally, since $G[T_i \cup T_j]$ contains at most two components, by (\ref{isolated-ei}) and by definition of paired sets one has that at most one from $\{Y_i,Y_j\}$ does exist.
$\diamond$ 

\medskip

The above claims are useful tools to detect $M_3$. Then let us observe that:

\begin{itemize}
\item[$(i)$] if a vertex $t_i \in T_i$ sees a vertex of $S_3 \cup S_4$, then $u_it_i \in M_3$;

\item[$(ii)$] if a vertex $t_i \in T_i$ is the center of the star $Y_i$ or $Y'_i$ (in case of paired sets), with a $P_3$ then $u_it_i \in M_3$.
\end{itemize}

Let us say that a vertex $t_i \in T_i$ is {\em green} if it enjoys one of the above two conditions $(i)$, $(ii)$.
Then the following statement holds for all $i = 1,\ldots,k$:

\begin{equation}\label{greenvertex1}
G[T_i] \mbox{ contains at most one green vertex, say } t_i^*
\end{equation}

and

\begin{equation}\label{greenvertex2}
G[T_i \setminus N(t_i^*)] \mbox{ is edgeless. }
\end{equation}

\section{Procedure Check$(xy)$}

In our algorithm $P_7$-Free-DIM in section \ref{generalP7freecase}, we carry out a fixed number of times the subsequent: 

\medskip

\noindent
{\bf Procedure Check$(xy)$}.

\medskip

\noindent
{\bf Given:} A (candidate) edge $xy$ which is in a $P_3$ of $G$.

\noindent
{\bf Task:} Determine a minimum weight d.i.m. $M$ of $G$ with $xy \in M$ or return a proof that $G$ has no d.i.m. with $xy$ or $G$ is not $P_7$-free.

\begin{enumerate}
\item[(a)] Determine the distance levels $N_1,N_2,\ldots$ with respect to $xy$.

\item[(b)] Check if all the conditions (\ref{NxysubI})-(\ref{isolated-ei}) of subsections \ref{levelstructure} and \ref{TiTjedges} are fulfilled. If one of them is not fulfilled then unsuccessfully STOP. 
    Otherwise, set $M: = \{xy\} \cup M_2 \cup M_4$. If $S_2 = \emptyset$, then successfully STOP - return $M$.

\item[(c)] Check if Claim \ref{edgesbetweenTiTj} of subsection \ref{TiTjedges} holds. If not, then unsuccessfully STOP. 
    Otherwise classify the $T_i$ sets into isolated ones and paired ones.

\item[(d)] Check if Claim \ref{eiTiTj} of subsection \ref{TiTjedges} holds. If not, then unsuccessfully STOP.

\item[(e)] Color green every vertex $t_i$ of $T_i$ such that either $t_i$ sees a vertex of $S_3 \cup S_4$ or $t_i$ is the center of the star $Y_i$ or $Y'_i$ (in case of paired sets) with $Y_i$ or $Y'_i$ containing $P_3$.

\item[(f)] Check if conditions (\ref{greenvertex1})-(\ref{greenvertex2}) of subsection \ref{TiTjedges} hold. If not, then unsuccessfully STOP. 

{\em Notation.} For any subset $T'_i$ of any $T_i$ set introduced in subsection \ref{TiTjedges}, let us say that a vertex $t'_i$ is a $best$ vertex in $T'_i$ if $w(u_it'_i) \leq w(u_it''_i)$ for any $t''_i \in T'_i$. $\diamond$

\item[(g)] For all isolated $T_i$, proceed as follows. If $T_i$ has a green vertex $t_i^*$, then set $M: = M
    \cup \{u_it^*_i\}$. Otherwise set $M: = M \cup \{u_it'_i\}$ where $t'_i$ is a best vertex in $Y_i$ (if $Y_i$ does exist) or is a best vertex in $T_i$ (otherwise).

\item[(h)] For all paired $T_i$ and $T_j$, proceed as follows.

\item[(h.1)] If $T_i$ and $T_j$ have a green vertex, respectively $t^*_i$ and $t^*_j$, then: if $t^*_i$ misses $t^*_j$, and if $G[(T_i \cup T_j) \setminus (N(t^*_i) \setminus N(t^*_j))]$ is edgeless then set $M: = M \cup \{u_it^*_i\} \cup \{u_jt^*_j\}$; otherwise unsuccessfully STOP. 

\item[(h.2)] If $T_i$ has a green vertex $t^*_i$, and if $T_j$ has no green vertex, then: If $G[(T_i \cup T_j) \setminus N(t^*_i)]$ has at least one vertex and contains most one component (i.e., $Y'_j$ or $Y_j$), then set $M: = M \cup \{u_it^*_i\} \cup \{u_jt_j\}$ where $t_j$ is, in this order, either the vertex in $Y'_j \cap T_j$ (if), or a best vertex in $Y_j$ (if), or a best vertex in $T_j$. Otherwise unsuccessfully STOP. If $T_j$ has a green vertex $t^*_j$, and if $T_i$ has no green vertex, then proceed similarly by symmetry.

\item[(h.3)] If $T_j$ and $T_j$ has no green vertex (according to Claim \ref{eiTiTj} and to the above, $G[T_i \cup T_j]$ contains isolated vertices, at most two isolated edges, and at least one isolated edge, say $t_it_j$, between $T_i$ and $T_j$), then proceed as follows:

\begin{itemize}
\item If there exists another edge, say $pq$, in $T_i$ or $T_j$ then: If $p,q \in T_i$ (or $p,q \in T_j$) then set $M := M \cup \{u_iz\} \cup \{u_jt_j\}$ where $z$ is a best vertex in $\{p,q\}$ (or $M := M \cup \{u_it_i\} \cup \{u_jz\}$ where $z$ is a best vertex in $\{p,q\}$); if $p \in T_i$ and $q \in T_j$, then either set $M := M \cup \{u_ip\} \cup \{u_jt_j\}$ or set $M := M \cup \{u_it_i\} \cup \{u_jq\}$, depending on the best alternative.

\item Otherwise: If $(T_i \setminus \{t_i\}) \cup (T_j \setminus \{t_j\}) = \emptyset$, then unsuccessfully STOP; if $T_i \setminus \{t_i\} \neq \emptyset$ and $T_j \setminus \{t_j\} = \emptyset$, then set $M := M \cup \{u_iz_i\} \cup \{u_jt_j\}$ where $z_i$ is a best vertex in $T_i \setminus \{t_i\}$; if $T_i \setminus \{t_i\} = \emptyset$ and $T_j \setminus \{t_j\} \neq \emptyset$, then set $M := M \cup \{u_it_i\} \cup \{u_jz_j\}$ where $z_j$ is a best vertex in $T_j \setminus \{t_j\}$; if $T_i \setminus \{t_i\} \neq \emptyset$ and $T_j \setminus \{t_j\} \neq \emptyset$, then either set $M := M \cup \{u_iz_i\} \cup \{u_jt_j\}$ where $z_i$ is a best vertex in $T_i \setminus \{t_i\}$, or set $M := M \cup \{u_it_i\} \cup \{u_jz_j\}$ where $z_j$ is a best vertex in $T_j \setminus \{t_j\}$, depending on the best alternative.
\end{itemize}

\item[(j)] Successfully STOP - return $M$.

\end{enumerate}

\begin{theorem}\label{maintheorem-procedure-xy-in-P3}
Procedure Check$(xy)$ is correct and runs in linear time.
\end{theorem}

\noindent
{\bf Proof.} {\em Correctness}: The correctness of the algorithm follows from the structural analysis of $P_7$-free graphs with d.i.m. described in subsections \ref{levelstructure} and \ref{TiTjedges}.

\medskip

\noindent
{\em Time bound}:
(a): Determining the distance levels $N_i$ with respect to edge $xy$ can be done in linear time, e.g. by using BFS.

\noindent
(b): Likewise, concerning conditions (\ref{NxysubI})-(\ref{isolated-ei}), we can test in linear time if $N_1$ is a stable set, $N_2$ is a disjoint union of edges and vertices, $N_5=\emptyset$, $N_4$ is a disjoint union of edges and vertices. The assignments can be done in linear time: This is obvious for $M, S_2$ and $S_4$. Then determine the degree of all vertices in $N_3$ with respect to $S_2$, and assign degree one vertices to $T_{one}$ and degree $\ge 2$ vertices to $T_{two}$. Obviously, a vertex in $N_3$ which misses $S_2$ has a predecessor in $M_2$, and thus $S_3$ and $T^*_{one}=T_{one} \setminus S_3$ form a partition of $N_3$. Obviously, it can be checked in linear time whether $N_3 \cup S_4$ is a bipartite subgraph and whether $S_3 \cup S_4$ is a stable set.

(c)-(j): All these steps can obviously be done in linear time.
\qed

\medskip

In the other case when an edge $xy$ is not in any $P_3$, it follows that $x$ and $y$ are true twins, and this case will be treated by determining the maximal homogeneous sets of $G$.   

\section{DIM for $P_7$-Free Bipartite Graphs}

A {\em domino} is a bipartite graph having six vertices, say $x_1,x_2,x_3,y_1,y_2,y_3$ such that $(x_1,y_1,x_2,y_2,x_3)$ is a $P_5$ and $y_3$ sees $x_1,x_2$ and $x_3$. 
 
\begin{observation}\label{obse:P7C6}
Let $M$ be a d.i.m. of a bipartite $P_7$-free graph $B$. 
\begin{itemize}
\item[$(i)$]
If $C$ is a $C_6$ in $B$ then exactly two $C$-edges are in $M$. 
\item[$(ii)$] $B$ is domino-free.  
\end{itemize}
\end{observation}

\noindent
{\bf Proof.} $(i)$: Assume to the contrary that the statement is not true. Let $C$ be a $C_6$ in $B$ with vertices $v_1,\ldots,v_6$ and edges $v_iv_{i+1}$, $i \in \{1,\ldots,6\}$ (index arithmetic modulo 6). Then by Observation \ref{dimC3C5C7C4} (iii), none of the $C$-edges are in $M$. Then since every edge of $B$ is matched by $M$, exactly three vertices of $C$, say $v_1,v_3,v_5$, belong to $V \setminus V(M)$, while $v_2,v_4,v_6$ belong to $V(M)$: let $v'_2,v'_4,v'_6$ be respectively their $M$-mates. Then by definition of $M$ and since $B$ is bipartite, $v'_2,v_2,v_3,v_4,v_5,v_6,v'_6$ induce a $P_7$ - contradiction. 

$(ii)$: If $D$ is a domino in $B$ then by Observation \ref{dimC3C5C7C4} (ii), the edges of the two $C_4$'s of $D$ must be matched from outside but now obviously there is a $P_7$ - contradiction.      
\qed

If moreover, $B$ is $C_6$-free, it is $(6,2)$-chordal bipartite, i.e., distance hereditary and bipartite (see e.g. \cite{BanMul1986}). In this case, DIM can be easily solved in linear time by using the clique-width argument \cite{CouMakRot2000,GolRot2000} since the clique-width of distance-hereditary graphs is at most three (and 3-expressions can be determined in linear time). We want to give a robust linear-time algorithm for $P_7$-free bipartite graphs for solving the DIM problem. If a bipartite graph $B$ is given, the algorithm either solves the DIM problem optimally or shows that there is a domino or $P_7$ in $B$. The algorithm constructs the distance levels starting from an arbitrarily chosen vertex. Then it checks whether $B$ is distance hereditary as in \cite{BanMul1986}. If a domino or $P_7$ is found, the algorithm unsuccessfully stops, and if a $C_6$ $C$ is found, one of the pairs of opposite edges in $C$ must be in $M$, say $v_1v_2$ and $v_4v_5$, and in this case, it is checked by Check($v_1v_2$) whether the distance levels starting from $v_1v_2$ have the required properties. 

\medskip

For making this paper self-contained, we repeat Corollary 5 of \cite{BanMul1986}:

\begin{corollary}[Bandelt, Mulder \cite{BanMul1986}]\label{bipdistheredneighb}
Let $G$ be a connected graph, and let $u$ be any vertex of $G$. Then $G$ is bipartite and distance hereditary if and only if all levels $N_k(u)$ are edgeless, and for any vertices $v,w \in N_k(u)$ and neighbors $x$ and $y$ of $v$ in $N_{k-1}(u)$, we have 
\begin{itemize}
\item[$(*)$] $N(x) \cap N_{k-2}(u) = N(y) \cap N_{k-2}(u)$, and further, 
\item[$(**)$] $N(v) \cap N_{k-1}(u)$ and $N(w) \cap N_{k-1}(u)$ are either disjoint, or one is contained in the other.  
\end{itemize}
\end{corollary}

We have to check level by level beginning with the largest index, whether $(*)$ and $(**)$ are fulfilled.   
If Condition $(*)$ is violated, we obtain a hole or domino. 
 
This leads to the following procedure for the bipartite case which includes a certifying recognition algorithm:  

\medskip

\noindent
{\bf Procedure $P_7$-Free-Bipartite-DIM}

\medskip

\noindent
{\bf Given:} A connected bipartite graph $B$ with edge weights.

\noindent
{\bf Task:} Determine a d.i.m. in $B$ of minimum weight (if existent) or find out that $B$ has no d.i.m. or is not $P_7$-free.

\begin{enumerate}
\item[(a)] Choose a vertex $a \in V$ and determine the distance levels $N_1,N_2,\ldots$ with respect to $a$. If $N_6 \neq \emptyset$ then STOP - $B$ is not $P_7$-free.

\item[(b)] For all levels $N_k$, $k \le 5$, beginning with $N_5$, check whether conditions $(*)$ and $(**)$ are fulfilled. If one of them is violated, we obtain an obstruction which is either a hole $C_8$ or $C_{10}$ (in the case of a $C_8$ or $C_{10}$ STOP - $B$ is not $P_7$-free), or a $C_6$ $C$ (in which case we have to proceed with $C$) or a domino - STOP - $B$ has no d.i.m. or is not $P_7$-free. 

\item[(c)] If in all cases, conditions $(*)$ and $(**)$ are fulfilled, $B$ is distance hereditary and bipartite. Apply the clique-width approach for solving the DIM problem. 

\item[(d)] ({\em Now $B$ is not distance hereditary and $C$ is a $C_6$ in $B$}.)  
For three consecutive edges $ab$ of $C$, carry out Check($ab$). If none of them ends successfully, STOP - $B$ has no d.i.m., otherwise we obtain an optimal d.i.m. (among the at most three solutions).   
\end{enumerate}

Check($ab$) assumes that $ab$ is in a $C_6$ of the bipartite graph $B$. In this case we have some additional properties, and the procedure could be simplified:

Let $N_{1a}=N(a) \cap N_1$ ($N_{1b}=N(b) \cap N_1$, respectively).
Obviously, the following is a partition of $N_1$ if $B$ is bipartite: 
\begin{equation}\label{N1partition}
N_1=N_{1a} \cup N_{1b}
\end{equation}

As before, $N_1$ has to be stable, and $N_2$ is a disjoint union of edges $M_2$ and vertices $S_2$. Since $ab$ is in a $C_6$, we have that 
$M_2 \neq \emptyset$. 

Since $B$ is $P_7$-free, obviously:
\begin{equation}\label{S2N4empty}
S_2 = \emptyset \mbox{ and } N_4 = \emptyset. 
\end{equation}  

Moreover:
\begin{equation}\label{N3stable}
N_3 \mbox{ is edgeless. } 
\end{equation}  

Finally, since $B$ is $P_7$-free, we obtain:
\begin{equation}\label{M2neighbinN1}
\mbox{Vertices in } M_2 \mbox{ of the same color have the same neighborhood in } N_1. 
\end{equation}

\noindent{\em Proof of $(\ref{M2neighbinN1})$}. Let $ef \in M_2$ and $gh \in M_2$ with $e$ and $g$ in the same color class, and suppose that $e$ sees $x \in N_{1a}$ while $g$ misses $x$. Then there is $y \in N_{1b}$ such that $yf \in E$. Since $N_1$ is stable, $xy \not\in E$. Since $g$ misses $x$, there is a neighbor $z \in N_{1a}$ of $g$. Since $h,g,z,a,x,e$ is no $P_7$, $ze \in E$. Again, since $N_1$ is stable, $yz \not\in E$. If $hy \in E$ then $x,e,z,g,h,y,b$ is a $P_7$. Thus, $hy \not\in E$ but now $h,g,z,a,b,y,f$ is a $P_7$ - a contradiction which shows (\ref{M2neighbinN1}). $\diamond$

\medskip

Obviously, $\{ab\} \cup M_2$ is a d.i.m. of $B$ if all conditions are fulfilled. 
 
\begin{lemma}\label{bipDIMcorr}
Procedure $P_7$-Free-Bipartite-DIM is correct and runs in linear time.
\end{lemma}

\noindent
{\bf Proof.} 
The correctness of the procedure follows from the structural analysis of bipartite $P_7$-free graphs with d.i.m. The time bound follows from the fact that procedure Check($xy$) is carried out only for a fixed number of edges, and each step of the procedure can be done in linear time.  
\qed

\section{Identifying an Odd Cycle in a Non-Bipartite $P_7$-Free Graph}

Let $G$ be a connected non-bipartite graph. The following procedure determines an odd cycle $C_3$, $C_5$ or $C_7$ or a $P_7$ of $G$ in linear time. 

\medskip

\noindent
{\bf Procedure Find-Odd-Cycle-Or-$P_7$}

\medskip

\noindent
{\bf Given:} A connected non-bipartite graph $G$.

\noindent
{\bf Task:} Determine an odd cycle $C_3$, $C_5$ or $C_7$ of $G$ or find out that $G$ is not $P_7$-free.

\begin{enumerate}
\item[(a)] Choose a vertex $x$ and determine the distance levels $N_1,N_2,\ldots$ with respect to $x$. If $N_6 \neq \emptyset$ then STOP - $G$ contains a $P_7$. 

\item[(b)] If there is an edge $ab \in E$ in $N_1$ then $xab$ is a $C_3$. Else $N_1$ is stable.

\item[(c)] If there is an edge $ab \in E$ in $N_2$ then $abc$ is a $C_3$ for a common neighbor $c \in N_1$ of $a,b$ or for neighbors $a' \in N_1$ of $a$ and $b' \in N_1$ of $b$, $xaba'b'$ is a $C_5$. Else $N_2$ is stable.

\item[(d)] If there is an edge $ab \in E$ in $N_3$ then $abc$ is a $C_3$ for a common neighbor $c \in N_2$ of $a,b$ or for neighbors $a' \in N_2$ of $a$ and $b' \in N_2$ of $b$, and a common neighbor $c \in N_1$ of $a',b'$, $caba'b'$ is a $C_5$ or for neighbors $a'' \in N_1$ of $a'$ and $b'' \in N_1$ of $b'$, $xa''b''a'b'ab$ is a $C_7$. Else $N_3$ is stable.

\item[(e)] If there is an edge $ab \in E$ in $N_4$ then $abc$ is a $C_3$ for a common neighbor $c \in N_3$ of $a,b$ or for neighbors $a' \in N_3$ of $a$ and $b' \in N_3$ of $b$, and a common neighbor $c \in N_2$ of $a',b'$, $caba'b'$ is a $C_5$ or for neighbors $a'' \in N_2$ of $a'$ and $a''' \in N_1$ of $a''$, $xa'''a''a'abb'$ is a $P_7$. Else $N_4$ is stable.

\item[(f)] ({\em Now $N_5$ must contain an edge, otherwise $G$ is bipartite.}) For an edge $ab$ in $N_5$, let $a_4$ denote a neighbor of $a$ in $N_4$ and let $a_{i-1} \in N_{i-1}$ denote a neighbor of $a_i \in N_i$, $i = 2,3,4$. Then either $a_4ab$ is a $C_3$ or $xa_1a_2a_3a_4ab$ is a $P_7$.    
\end{enumerate}
 
Obviously, the following holds: 
\begin{lemma}\label{findoddcyccorr}
Procedure Find-Odd-Cycle-Or-$P_7$ is correct and runs in linear time.
\end{lemma}
 
\section{The Algorithm for the General $P_7$-Free Case}\label{generalP7freecase}

In the previous chapters we have analyzed the structure of $P_7$-free graphs having a d.i.m. Now we are going to use these properties for an efficient algorithm for solving the DIM problem on these graphs. 
\medskip  

\noindent
{\bf Algorithm $P_7$-Free-DIM.}

\medskip

\noindent
{\bf Given:} A connected graph $G=(V,E)$ with edge weights.

\noindent
{\bf Task:} Determine a d.i.m. in $G$ of finite minimum weight (if existent) or find out that $G$ has no d.i.m. or is not $P_7$-free.

\begin{enumerate}
\item[(a)] If $G$ is bipartite then carry out procedure $P_7$-Free-Bipartite-DIM. 

\item[(b)] ({\em Now $G$ is not bipartite.}) If $G$ is a cograph then apply procedure Cograph-DIM. If $G$ is not a cograph but $\overline{G}$ is not connected then STOP - $G$ has no d.i.m. 

\item[(c)] ({\em Now $G$ is not bipartite and $\overline{G}$ is connected.}) Let $M:=\emptyset$. 
Determine the maximal homogeneous sets $H_1,\ldots,H_k$ of $G$. For all $i \in \{1,\ldots,k\}$ do the following steps (c.1), (c.2):

\item[(c.1)] If $|N(H_i)| = 1$ then carry out procedure Hom-1-DIM. 

\item[(c.2)] In the case when $|N(H_i)| \ge 2$ and $H_i$ is not a stable set, check whether $N(H_i)$ is stable and $H_i$ is a disjoint union of edges; if not then STOP - $G$ has no d.i.m., otherwise, for all edges $xy$ in $H_i$, let $M:= M \cup \{xy\}$. 

\item[(d)] Construct $G'=Reduced(G,M)$.

\item[(e)] For every connected component $C$ of $G'$, do: If $C$ is bipartite then carry out procedure $P_7$-Free-Bipartite-DIM for $C$.  
Otherwise construct $C^*$ (where the triangle leaf blocks are temporarily omitted) and carry out Find-Odd-Cycle-Or-$P_7$ for $C^*$, and if an odd cycle is found, carry out Check($ab$) in the graph $C$ for all (at most seven) edges of the odd cycle. Add the resulting edge set to the mandatory edges from steps (c.1), (c.2), respectively. If however, $C^*$ is bipartite then with procedure $P_7$-Free-Bipartite-DIM for $C^*$, find out if the procedure unsuccessfully stops or if there is a $C_6$ in $C^*$; in the last case, do Check($ab$) for all edges of the $C_6$. Finally, if  $C^*$ is distance hereditary bipartite, construct $Tr(C)$  (the omitted triangle leaf blocks are attached as $P_3$'s and the resulting graph is distance hereditary bipartite) and solve DIM for this graph using the clique-width argument (or using the linear time algorithm for chordal bipartite graphs given in \cite{BraHunNev2010}).      

\item[(f)] Finally check once more whether $M$ is a d.i.m. of $G$. If not then $G$ has no d.i.m., otherwise return $M$.
\end{enumerate}

\begin{theorem}\label{DIMP5theorem}
Algorithm $P_7$-Free-DIM is correct and runs in linear time.
\end{theorem}

\noindent
{\bf Proof.}
{\em Correctness}:
The correctness of the algorithm follows from the structural analysis of $P_7$-free graphs with d.i.m. In particular, if $G$ is bipartite (a cograph, respectively) then procedure $P_7$-Free-Bipartite-DIM (Cograph-DIM, respectively) correctly solves the DIM problem. 

If $\overline{G}$ is not connected, i.e., $G = G_1 \join G_2$ for some nonempty $G_1,G_2$ and $G$ has a d.i.m. then by Corollary \ref{cor:coconnwithdim}, $G$ must be a cograph. 

For the maximal homogeneous sets $H_1,\ldots,H_k$ of $G$, there are two cases $|N(H_i)| = 1$ or $|N(H_i)| \ge 2$. By Corollary \ref{homsets} and Lemma \ref{Hom1DIMcorrlintime}, steps (c.1) and (c.2) are correct, and $G$ can be correctly reduced by using the obtained set $M$ of forced edges. Since in procedure Hom-1-DIM, in the last two cases, the corresponding leaf blocks are postponed, in the reduced graph, every odd cycle contains only edges in $P_3$'s. Thus, it is correct to apply Check($ab$) for the edges of some odd cycle in the (non-bipartite) reduced graph. Finally one has to add the postponed edges and solve the DIM problem on these graphs.     

\medskip

\noindent
{\em Time bound}: Step (a) can be done in linear time since procedure $P_7$-Free-Bipartite-DIM takes only linear time.
Step (b) can be done in linear time since it can be recognized in linear time whether $G$ is a cograph (see \cite{BreCorHabPau2008,CorPerSte1985}) and procedure Cograph-DIM can be done in linear time.
Step (c) can be done in linear time since modular decomposition can be done in linear time and gives the maximal homogeneous sets \cite{McCSpi1999}. There is only a linear number of true twins, and the corresponding reduced graph can be determined in linear time. 
 
In the reduced graph $G'=Reduced(G,M)$, procedure Check($xy$) is carried out only for a fixed number of edges, and the procedures  $P_7$-Free-Bipartite-DIM and Find-Odd-Cycle-Or-$P_7$ can be done in linear time.
\qed

\section{Conclusion}

In this paper we solve the DIM problem in linear time for $P_7$-free graphs which answers an open question from \cite{CarKorLoz2011}.
Actually, we solve the minimum weight DIM problem in a robust way in the sense of \cite{Spinr2003}: Our algorithm either solves the problem correctly or finds out that the input graph has no d.i.m. or is not $P_7$-free. This avoids to recognize whether the input graph is $P_7$-free; the known recognition time bound is much worse than linear time.

It is a challenging open question whether for some $k$, the DIM problem is \NP-complete for $P_k$-free graphs.

\medskip

\noindent
{\bf Acknowledgement.}
The first author gratefully acknowledges a research stay at the LIMOS institute, University of Clermont-Ferrand, and the inspiring discussions with Anne Berry on dominating induced matchings. 

\begin{footnotesize}

\end{footnotesize}

\end{document}